\documentclass[12pt]{article}

\newcommand{\bea}{\begin{eqnarray}}
\newcommand{\eea}{\end{eqnarray}}
\newcommand{\be}{\begin{equation}}
\newcommand{\ee}{\end{equation}}

\usepackage{amsmath}
\usepackage{bm}
\usepackage[dvips]{graphicx}
\baselineskip
\begin{document}

\begin{flushright}
nucl-th/0310048\\
\end{flushright}

\begin{center}
\vspace*{0cm}

{\large {\bf Renormalization of EFT for nucleon-nucleon scattering}} \\

\vspace*{1cm}

J.-F. Yang$^{1}$

\vspace*{0.2cm}
$^{1}${\small Department of Physics,\\
East China Normal University,\\
Shanghai 2000062, China.}

\end{center}
\vspace*{.5cm}
\begin{abstract}{
The renormalization of EFT for nucleon-nucleon scattering in
nonperturbative regimes is investigated in a compact
parametrization of the $T$-matrix. The key difference between
perturbative and nonperturbative renormalization is clarified. The
underlying theory perspective and the 'fixing' of the
prescriptions for the $T$-matrix from physical boundary conditions
are stressed.}
\end{abstract}


\setcounter{equation}{0}
\section{}
In recent years, the effective field theory (EFT) method has
become the main tool to deal with various low energy processes and
strong interactions in the nonperturbative regime. Beginning from
Weinberg's seminal works\cite{WeinEFT}, this method has been
extensively and successfully applied to the low energy nucleon
systems\cite{BevK}. As an EFT parametrizes the high energy details
in a simple way, there appear severe UV divergences (or
ill-definedness) and more undetermined constants that must be
fixed somehow. Owing to the nonperturbative nature, the EFT for
nuclear forces also becomes a theoretical laboratory for studying
the nonperturbative renormalization and a variety of
renormalization (and power counting) schemes have been
proposed\cite{vK,Epel,Cohen,Rho,KSW,Gege,Soto,BBSvK,Oller,Nieves,VA}\footnote{
We apologize in advance for those contributions we have not been
informed of by now.}. However, a complete consensus and
understanding on this important issue has yet not been reached.
Though the EFT philosophy is very natural, but its implementation
requires discretion (in parametrizing the short distance
effects)\cite{Burgess}. In particular, as will be shown below, the
nonperturbative context impedes the implementation of conventional
subtraction renormalization, and some
regularization/renormalization (R/R) prescription might even fail
the EFT method or physical predictions\cite{Cohen,Soto,YangPRD02},
unlike in the perturbative case\cite{scheme}.

In this short report, we shall revisit the problem without
specifying the R/R scheme (we only need that the renormalization
is done) to elucidate the key features for  the implementation of
R/R in nonperturbative regime, and various proposals could be
compared and discussed. Then the key principles that should be
followed are suggested as the first steps in such understandings.

\section{}
The physical object we are going to investigate is the transition
matrix $T$ for nucleon-nucleon scattering processes at low
energies,
\begin{equation} \label{NN} N({\bf p})+N({\bf -p})\rightarrow
N({\bf p}^{\prime})+N({\bf -p}^{\prime}),\end{equation}which
satisfies the Lippmann-Schwinger (LS) equation in partial wave
formalism,
\begin{eqnarray}
\label{LSE} &&T_{ll^\prime}(p^{\prime},p;
E)=V_{ll^\prime}(p^{\prime},p)+\sum_{l^{\prime\prime}}\int
\displaystyle\frac{kdk^2}{(2\pi)^2} V_{l
l^{\prime\prime}}(p^{\prime},k ) G_0(k;E^+)
T_{l^{\prime\prime}l^\prime}(k,p; E), \nonumber \\
&&G_0(k;E^+)\equiv \frac{1}{E^{+}-k^2/(2\mu)},\ \ E^{+}\equiv E +
i \epsilon,
\end{eqnarray}
with $E$ and $\mu$ being respectively the center energy and the
reduced mass, ${\bf p}$ (${\bf p}^\prime$) being the momentum
vector for the incoming (outgoing) nucleon, and $p^{\prime}=|{\bf
p}^\prime|$, $p=|{\bf p}|$. The potential $V(p^{\prime},p)$ can be
systematically constructed from the $\chi$PT\cite{ChPT,WeinEFT}.
We remind that the constructed potential is understood to be
finite first, as 'tree' vertices in the usual field theory
terminology.

Now let us transform the above LS equation into the following
compact form as a nonperturbative parametrization of $T$-matrix,
we temporarily drop all the subscripts that labelling the status
of the finiteness and divergence:
\begin{eqnarray}
\label{npt} &&T^{-1}=V^{-1}-{\mathcal{G}},
\end{eqnarray}where ${\mathcal{G}}\equiv
V^{-1}\{\int \frac{kdk^2}{(2\pi)^2}V G_0T\} T^{-1}$ is obviously a
nonperturbative quantity. Here $V$, $T$ and therefore
${\mathcal{G}}$ are momenta dependent matrix in angular momentum
space. In field theory language, ${\mathcal{G}}$ contains all the
'loop' processes generated by the 4-nucleon vertex $V$ joined by
nucleon (and antinucleon) lines\cite{KSW}. This nonperturbative
parametrization clearly separates $V$ (Born amplitude) from
${\mathcal{G}}$ (all the iterations or rescatterings), and
on-shell ($p^{\prime}=p=\sqrt{2\mu E}$) unitarity (nonperturbative
relation) of $T$ is automatically satisfied here thanks to the
on-shell relation between the $K$-matrix and
$T$-matrix\cite{newton}
\begin{eqnarray}\label{KT}T^{-1}_{os}=K^{-1}_{os}+i\mu \frac{\sqrt{2\mu
E}}{2\pi},\end{eqnarray} Using Eq.(~\ref{npt}) we have
\begin{eqnarray}{\mathcal{G}}_{os}=V_{os}^{-1}-K_{os}^{-1}-i\mu
\frac{\sqrt{2\mu E}}{2\pi}.\end{eqnarray} Any approximation to the
quantity ${\mathcal{G}}$ is leads to a nonperturbative scheme for
$T$.

Since the on-shell $T$-matrix must be physical, then for general
potential functions this nonperturbative parametrization leads to
the following obvious but important observations in turn: (I) it
is impossible to renormalize ${\mathcal{G}}$ through conventional
perturbative counter terms in $V$ (we will demonstrate this point
shortly), i.e., they have to be separately renormalized; (II) the
perturbative pattern of the scheme dependence for transition
matrix or physical observables\cite{scheme} breaks down in the
nonperturbative contexts like Eq.(~\ref{npt}); (III) the R/R for
$V$ and ${\mathcal{G}}$ must be so performed that the on-shell
$T$-matrix satisfies physical boundary conditions (bound state
and/or resonance poles\cite{Jackiw,KSW}, scattering lengths or
phase shifts\cite{nij}, etc.), i.e, $V$ and ${\mathcal{G}}$ are
'locked' with each other; (IV) due to this 'locking', the validity
of the EFT power counting in constructing $V$ also depends on the
R/R prescription for ${\mathcal{G}}$, i.e., the power counting of
$V$ could be vitiated by the R/R prescription for ${\mathcal{G}}$
which is performed afterwards. This explains why there are a lot
of debates on the consistent power counting schemes for
constructing $V$. Point (III) is in fact what most authors
actually did or is implicit in their treatments. we will return to
these points later.

To be more concrete or to demonstrate the first point (observation
(I)), let us consider the case without mixing ($l=l^{\prime}$) for
simplicity. Then the matrices become diagonal and only energy
dependent due to on-shell condition, we have the following simple
form of nonperturbative parametrization (from now on we will drop
the subscript "os" for on-shell)\footnote{We can also arrive at it
simply through dividing both sides of Eq.(~\ref{LSE}) by
$V_{l}(E)$ and $T_{l}(E)$ and rearranging the terms.},
\begin{eqnarray}
\label{partialW}
\frac{1}{T_l(E)}=\frac{1}{V_l(E)}-{\mathcal{G}}_l(E),
\end{eqnarray}where ${\mathcal{G}}_{l}(E)=\frac{(\int
\frac{kdk^2}{(2\pi)^2}V G_0T)_{l}(E)}{V_{l}(E)T_{l}(E)}$. In
perturbative formulation, all the divergences are removed order by
order {\em before} the loop diagrams are summed up. Here, in
Eq.(~\ref{partialW}) (or Eq.(~\ref{npt})), the renormalization are
desired to be performed on the compact nonperturbative
parametrization. After $V$ is calculated and renormalized within
$\chi$PT, the task is (a) to remove the divergence or
ill-definedness in ${\mathcal{G}}_l(E)$ and (b) to make sure that
the renormalized quantities ($V^{(R)}_l (E)$ and
${\mathcal{G}}^{(R)}_l(E)$) lead to a physical $T$-matrix. The
most important one is (b), namely the $T$-matrix obtained must
possess reasonable or desirable physical behaviors after removing
the divergences. Technically, this amounts to a stringent
criterion for the R/R prescription in use: the functional form of
${\mathcal{G}}_l(E)$ could not be altered, only the divergent
parameters (coefficients) get 'replaced' by finite ones.

Now let us suppose ${\mathcal{G}}_l$ could be renormalized by
introducing counter terms. Within this parametrization of
Eq.(~\ref{partialW}), the potential $V$ appears as the only
candidate to bear counter terms, then from Eq.(~\ref{partialW}) we
would have
\begin{eqnarray}
\label{ct}{\mathcal{G}}^{(R)}_l(E)={\mathcal{G}}^{(B)}_l(E)+
\frac{1}{V^{(R)}_l(E)}- \frac{1}{V^{(R)}_l(E)+\delta V_l (E)},
\end{eqnarray}where the superscripts $(R)$ and $(B)$ refer to
'renormalized' and 'bare', $\delta V_l$ denotes the additive
counter terms and $V^{(B)}_l (E)\equiv V^{(R)}_l(E)+\delta V_l
(E)$.

From the definition below Eq.(~\ref{npt}), ${\mathcal{G}}$ must
take the following form which is no less complicated than a
nonpolynomial function in $p(=\sqrt{2\mu E})$,
\begin{equation}
{\mathcal{G}}^{(B)}_l(E)=\frac{\int \frac{d^3 k}{(2\pi)^3} V_l(p,k
) G_0(k;E^+) T_l(k,p; E)}{ V_l(E)
T_l(E)}=\frac{N^{(B)}_l(p)}{D^{(B)}_l(p)}- \frac{\mu}{2\pi}i p,
\end{equation}with $N^{(B)}_l(p),D^{(B)}_l(p)$ being at least two
nontrivial divergent or ill-defined polynomials. With this
parametrization, Eq.(~\ref{ct}) becomes
\begin{equation}
\label{ct2} {\mathcal{G}}^{(R)}_l(E)=\frac{N^{(B)}_l(p)
V^{(R)}_l(E) V^{(B)}_l(E)+D^{(B)}_l(p) \delta V_l (E)
}{D^{(B)}_l(p)V^{(R)}_l(E) V^{(B)}_l(E)}- \frac{\mu}{2\pi}i p
=\frac{N^{(R)}_l(p)}{D^{(R)}_l(p)}- \frac{\mu}{2\pi}i p.
\end{equation} Note that in the divergent or ill-defined fraction
$\frac{N^{(B)}_l(p) V^{(R)}_l(E) V^{(B)}_l(E)+D^{(B)}_l(p) \delta
V_l (E) }{D^{(B)}_l(p)V^{(R)}_l(E) V^{(B)}_l(E)}$ only
$V^{(R)}_l(E)$ is finite. Then except for the simplest form of
$V_l (E)$ (= $C_0$, see, e.g., Ref.\cite{BevK}), it is impossible
to obtain a nontrivial and finite fraction out of the this
divergent fraction with {\em whatever} counter terms $\delta V_l
(E)$, i.e., it is impossible to remove the divergences in the
numerator and the denominator at the same time {\em with the same}
counter terms as in Eq.(~\ref{ct2}). Examples for nonperturbative
divergent fractions could be found in Ref.\cite{Cohen}. Even by
chance ${\mathcal{G}}$ is finite, then its functional form in
terms of physical parameters ($E,p$) must have been altered (often
oversimplified) after letting the cutoffs go infinite. For
channels with mixing, with $V_l$ and ${\mathcal{G}}_l$ becoming
$2\times2$ matrices, all the preceding deductions still apply.
Thus within nonperturbative regime the counter term (via
potential) renormalization of $T$-matrix fails, that is,
${\mathcal{G}}$ must be separately renormalized--the first
observation given above is demonstrated as promised. The other
observations follow easily.

What we have just shown does not mean that the counter term fails
at any rate. The above arguments showed that the counter term
could not successfully renormalize the nonperturbative
$\mathcal{G}$ or $T$-matrix {\em except} they are introduced {\em
before} the corresponding infinite perturbative series are summed
up or {\em before} all parts of the nonperturbative objects are
put together and 'installed', say, before the corresponding
Schr\"odinger equation is solved. It is well known that the
Schr\"odinger equation approach\cite{BBSvK} or similar\cite{VA} is
intrinsically nonperturbative, therefore the counter terms
introduced there (as parts of the potential operators) ARE
nonperturbative in the sense that they enter {\em before} the
$T$-matrix or phase shifts are finally derived, or equivalently,
they must 'enter' before the perturbative series are summed up. So
there is no contradiction between our conclusion here and that in
Ref.\cite{BBSvK}. In fact they are in complete harmony and our
conclusion above turns out to be also supporting the Schr\"odinger
equation approach for renormalizing nucleon-nucleon scattering.
Thus our discussions above are helpful to clarify the difference
between perturbative and nonperturbative renormalization or to
remove some mists or controversies around the renormalization of
EFT for nucleon-nucleon scattering.

To make our language more concise, we would temporarily call the
counter terms 'endogenous' to an object if they must enter before
all the necessary calculations on the interested object are done,
and 'exogenous' otherwise.

\section{}
Therefore the renormalization in nonperturbative regime have to be
implemented either with 'endogenous' counter terms
(e.g.,\cite{BBSvK}) or otherwise (e.g., like in
Refs.\cite{Nieves,VA}). To proceed we recall the conceptual
foundation for EFT method: there must be a complete theory
underlying all the low energy EFTs. Suppose we could compute the
$T$-matrix in the underlying theory, then in the low energy limit,
we should obtain a finite matrix element parametrized in terms of
EFT coupling constants and certain {\em finite constants arising
from the limit}. It is these constants that we are after. In the
ill-defined EFT frameworks, a R/R or subtraction prescription is
employed to 'retrieve' the finite constants. As we have seen the
failure of the counter term subtraction at the potential or vertex
level in nonperturbative parametrization, the renormalization must
be otherwise implemented. A prompt choice is the subtraction at
the integral level, and the counter term formalism is dismissed
from the underlying theory point of view. In this sense, the
formal consistency issue of Weinberg power
countings\cite{KSW,BevK} is naturally dissolved without resorting
to other means, and the real concern is the renormalization
instead of power counting rules. This resolution is in fact just a
paraphrase of the fourth observation in section two (point (IV)):
a consistent power counting could be vitiated by the afterward
renormalization procedures such as those using exogenous counter
terms.

If the ill-definedness only originates from the local part of the
potential, the integral level subtraction is OK as the divergent
integrals can be cleanly isolated\cite{Gege,Nieves}\footnote{For
more general parametrization of the ill-defined integrals,
see\cite{PQFT98,YangPRD02}.}. For the more interesting cases with
general pion-exchange contributions, such clean isolation seems
impossible, which leads some authors to perturbative approach
(KSW\cite{KSW}), but the convergence is shown to be
slow\cite{KSW,BevK} and the KSW power counting also bears some
problems\cite{BBSvK}. Due to such difficulties, the regularization
schemes with finite cutoffs or parameters have also been adopted
to keep the investigations
nonperturbative\cite{vK,Epel,Rho,BBSvK}\footnote{All the finite
cutoff regularization schemes could be seen as effectively
employing endogenous counter terms implicitly: the UV region
contributions are subtracted before integration.}, which have to
be numerical. In such approaches certain sort of 'fitting to data'
is incorporated in one way or the other, this is what we stressed
in section two. There recently appears a new approach that
constructs the nonperturbative $T$-matrix from the KSW
approach\cite{Oller}, where again certain sort of fitting is
crucial there.

Then it is clear that the main obstacle is the severe
regularization dependence in nonperturbative regime. If we could
parametrize all the ill-definedness in ambiguous but finite
expressions, then the obstacle would disappear and the only task
is to fix the ambiguities. Otherwise, any theoretical judgement
just based on one regularization scheme is unwarranted or
ungrounded unless it is justified in various schemes
\cite{Soto,chLnforce}. The importance of regularization in
nonperturbative regime could also be seen in the following way:
For any R/R prescription in nonperturbative regimes to be able to
'reproduce' such 'true' $T$-matrix obtained from underlying
theory, the regularization in use should at least reproduce the
same nonperturbative functional expression as that given in
underlying theory, with certain constants being adjustable,
otherwise the 'true' functional expression could not be approached
since exogenous counter terms, as we have shown above, could not
be effected in nonperturbative parametrization. Alternatively, the
foregoing arguments could be interpreted as a call for a devise of
a new nonperturbative framework where endogenous counter terms
could be naturally incorporated.

\begin{figure}[h]
\begin{center}
\includegraphics*[width=10cm]{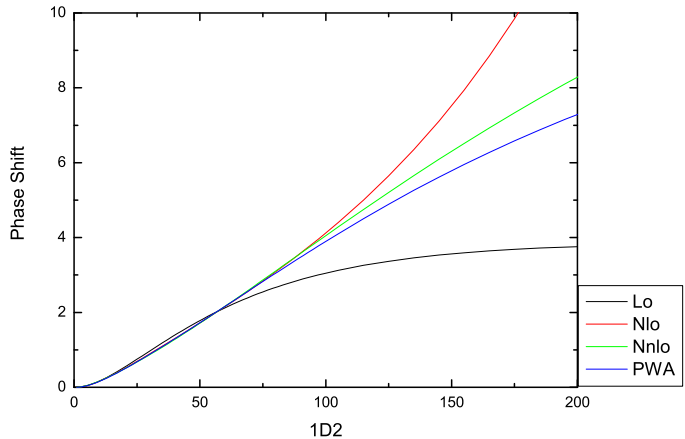}
\caption{\small The prediction of the phase shifts versus energy
$E$ (Mev) of $^1D_2$ channel at LO, Nlo and Nnlo with the simplest
parametrization $Re({\mathcal{G}}_{l=2})=g_{l=2;0}$. Also listed
is the curve for PWA. For each order the $g_{2;0}$ is fitted to or
optimized in the low energy region of the PWA data.} \label{1d2}
\end{center}
\end{figure}

Our parametrization also points towards a new technical direction
for the nonperturbative determination of the $T$-matrix: to focus
on the calculation of ${\mathcal{G}}_l(E)$. The relevance of R/R
in ${\mathcal{G}}$ can be made clearer in the low energy expansion
of ${\mathcal{G}}$ (Pad\'e or Taylor) since it is necessarily a
function of the center energy ($E$) or on-shell momentum ($p$):
\begin{equation}
\label{PADE} Re({\mathcal{G}}_l (p))|_{\texttt{Pad\'e}}=
\frac{n_{l;0}+n_{l;1}p^2+\cdots}{d_{l;0}+d_{l;1}p^2
+\cdots}|_{\texttt{Taylor}} = g_{l;0} +g_{l;1} p^2 +g_{l;2} p^4
+\cdots. \end{equation} Here the coefficients $ g_{l;n }$ will
inevitably be R/R dependent and must be determined together with
the parameters in the potential from physical boundary conditions,
say the empirical phase shifts in the low energy ends\cite{nij}.
Investigations along this line are in progress\cite{YH} and the
primary the results are interesting. In fact, there is several
virtues in such analysis: Firstly, one needs not to carry out the
concrete the renormalization of ${\mathcal{G}}$; Secondly, by
comparing with low energy region of the empirical data of $NN$
scattering (say, $E\leq 50 \texttt{Mev}$) one could in principle
easily obtain the optimal values for the semi-phenomenological
parameters ($g_{l;n }$) which in turn could yield a pretty good
prediction of higher energy behaviors though the parametrization
formula of Eq.(~\ref{npt}) is strikingly simple; Thirdly, with
such simple tools one could test whether EFT systematically works
for $NN$ scattering; Fourthly, one could also use it to estimate
if the potential constructed up to a certain order is sufficient
for various purposes {\em without any loop calculations and
renormalization operations}. Moreover, due to its simplicity in
principle, one might also find other uses like the estimation of
the coupling constants of the local terms in constructed
potential. This list could be further enlarged. We will visit them
in the near future. In Fig.~\ref{1d2} we serve a primary example
for such analysis\cite{YH} employing the potential given by
EGM\cite{Epel}: the phase shifts of $^1D_2$ channel predicted only
with a {\em single} parameter $Re({\mathcal{G}}_{l=2})\approx
g_{l=2;0}$ with potentials input at leading order (Lo),
next-to-leading order (Nlo) and next-to-next-to-leading order
(Nnlo), respectively. It is clear that the prediction of the phase
shift for the range $50\texttt{Mev} \le E \le 200\texttt{Mev}$
with only one optimized or fitted parameter of
$Re({\mathcal{G}}_{l=2})$, $g_{l=2;0}$, improves with chiral order
for the construction of the potential (Lo, Nlo, and Nnlo). For
most channels this seems true. Of course, there could well be
exceptions, which we interpret it as insufficiency of the chiral
expansion in the corresponding channel(s).

In summary, we proposed a simple and novel parametrization for
understanding the nonperturbative renormalization of the
$T$-matrix for nucleon-nucleon scattering. The distinctive feature
of the nonperturbative renormalization--the failure of 'exogenous'
counter terms or the need of 'endogenous' ones is clarified
together with some related issues like consistency of W power
counting. By the way we could derive other uses from our simple
parametrization of the $T$-matrix that might be of some academic
values. We hope our investigation might be useful for a number of
important issues related to nucleon interactions, and also to
other nonperturbative problems.

\section*{Acknowledgement}
I wish to thank Dr. E. Ruiz Arriola and Dr. J. Gegelia for helpful
communications on the topic. I am especially grateful to an
anonymous referee for his very helpful and encouraging report that
leads to the present form of this manuscript. This project is
supported in part by the National Natural Science Foundation under
Grant No. 10502004.


\begin{thebibliography}{99}
\bibitem{WeinEFT} S. Weinberg, Phys. Lett. {\bf B 251}, 288 (1990);
Nucl. Phys. {\bf B 363}, 1 (1991).
\bibitem{BevK} See, e.g., P. Bedaque and U. van
Kolck, Ann. Rev. Nucl. Part. Sci. \textbf{52}, 339 (2002).
\bibitem{vK}C. Ord\'o\~nez, L. Ray and U. van Kolck,
Phys. Rev. \textbf{C53}, 2086 (1996); U. van Kolck, Nucl. Phys.
\textbf{A645}, 327 (1999).
\bibitem{Epel}E. Epelbaum, W. Gl\"ockle and U. Meissner, Nucl. Phys.
\textbf{A637}, 107 (1998), Nucl. Phys. \textbf{A671}, 295 (2000),
Eur. Phys. J. \textbf{A15}, 543 (2002), arXiv:nucl-th/0304037,
0308010.
\bibitem{Cohen} D.R. Phillips, S.R. Beane and T.D.
Cohen, Ann. Phys.(NY) \textbf{263}, 255(1998) and references
therein.
\bibitem{Rho} J.V. Steele and R.J. Furnstahl, Nucl. Phys.
\textbf{A637}, 46 (1999);  T.S. Park, K. Kubodera, D.P. Min and M.
Rho, Phys. Rev. \textbf{C58}, 637 (1998); T. Frederico, V.S.
Tim\'oteo and L. Tomio, Nucl. Phys. \textbf{A653}, 209 (1999).
\bibitem{KSW} D.B. Kaplan, M.J. Savage and M.B. Wise,
Phys. Lett. \textbf{B424}, 390 (1998); Nucl. Phys. \textbf{B534},
329 (1998); S. Fleming, T. Mehen and I.W. Stewart, Nucl. Phys.
\textbf{A677}, 313 (2000); Phys. Rev. \textbf{C61}, 044005 (2000).
\bibitem{Gege} J. Gegelia, Phys. Lett. \textbf{B463}, 133 (1999);
J. Gegelia and G. Japaridze, Phys. Lett. \textbf{B517}, 476
(2001); J. Gegelia and S. Scherer, arXiv:nucl-th/0403052.
\bibitem{Soto} D. Eiras and J. Soto, Eur. Phys.
J. \textbf{A17}, 89 (2003).
\bibitem{BBSvK} S.R. Beane, P. Bedaque, M.J. Savage and U. van Kolck, Nucl.
Phys. \textbf{A700}, 377 (2002).
\bibitem{Oller} J.A. Oller, arXiv:nucl-th/0207086.
\bibitem{Nieves} D.B. Kaplan, Nucl. Phys. \textbf{B494}, 471
(1997); J. Nieves, Phys. Lett. \textbf{B568}, 109 (2003).
\bibitem{VA} M. Pavon Valderrama and E. Ruiz
Arriola, Phys, Lett. \textbf{B580}, 149 (2004),
arXiv:nucl-th/0405057.
\bibitem{Burgess} C.P. Burgess and D. London, Phys. Rev.
\textbf{D48}, 4337 (1993).
\bibitem{YangPRD02} For examples in other contexts, see, e.g., J.-F. Yang
and J.-H. Ruan, Phys. Rev. \textbf{D65}, 125009 (2002) and
references therein.
\bibitem{scheme}  P.M. Stevenson, Phys. Rev. \textbf{D23}, 2916 (1981);
G. Grunberg, Phys. Rev. \textbf{D29}, 2315 (1984); S.J. Brodsky,
G.P. Lepage and P.B. Mackenzie, Phys. Rev. \textbf{D28}, 228
(1983).
\bibitem{ChPT}S. Weinberg, Physica \textbf{A96}, 327 (1979); J.
Gasser and H. Leutwyler, Ann. Phys. \textbf{158}, 142 (1984); J.
Gasser and H. Leutwyler, Nucl. Phys. \textbf{B250}, 465 (1985).
\bibitem{newton} R.G. Newton, {\em Scattering Theory of Waves and
Particles}, 2nd Edition (Springer-Verlag, New York, 1982), p187.
\bibitem{Jackiw} R. Jackiw, in {\it M.A.B. B\' eg Memorial
Volume}, A. Ali and P. Hoodbhoy, eds. ( World Scientific,
Singapore, 1991), p25-42.
\bibitem{nij} V.G.J. Stoks, R.A.M. Klomp, M.C.M. Rentmeester, and
J.J. de Swart, Phys. Rev. \textbf{C48}, 792 (1993).
\bibitem{PQFT98} Ji-Feng Yang, arXiv: hep-th/9708104; invited
talk in: {\em Proceedings of the XIth International Conference
'Problems of Quantum Field Theory'98'}, Eds. B. M. Barbashov {\em
et al}, (Publishing Department of JINR, Dubna, 1999),
p.202[arXiv:hep-th/9901138]; arXiv: hep-th/9904055.
\bibitem{chLnforce} See the recent debates on the chiral limit nuclear
forces S.R. Beane and M.J. Savage, Nucl. Phys. \textbf{A713}, 148
(2003); Nucl. Phys. \textbf{A717}, 91 (2003); E. Epelbaum, W.
Gl\"ockle and U.-G. Meissner, Nucl. Phys. \textbf{A714}, 535
(2003).
\bibitem{YH} J.-F. Yang and Jian-Hua Huang, in preparation.
\end{thebibliography}
\end{document}